# A non-intrusive measurement technique applying CARS for concentration measurement in a gas mixing flow

Ken Yamamoto, Yuki Yamagata, Madoka Moriya, Reiko Kuriyama and Yohei Sato

Department of System Design Engineering, Faculty of Science and Technology, Keio University, 3-14-1 Hiyoshi, Kohoku-ku, Yokohama 223-8522, Japan

Email: ken-yamamoto@tfe.sd.keio.ac.jp

**Abstract**

Coherent anti-Stokes Raman scattering (CARS) microscope system was built and applied to a non-intrusive gas concentration measurement of a mixing flow in a millimeter-scale channel. Carbon dioxide and nitrogen were chosen as test fluids and CARS signals from the fluids were generated by adjusting the wavelengths of the Pump and the Stokes beams. The generated CARS signals, whose wavelengths are different from those of the Pump and the Stokes beams, were captured by an EM-CCD camera after filtering out the excitation beams. A calibration experiment was performed in order to confirm the applicability of the built-up CARS system by measuring the intensity of the CARS signal from known concentrations of the samples. After confirming that the measured CARS intensity was proportional to the second power of the concentrations as was theoretically predicted, the CARS intensities in the gas mixing flow channel were measured. Ten different measurement points were set and concentrations of both carbon dioxide and nitrogen at each point were obtained. Consequently, it was observed that the mixing of two fluids progressed as the measurement point moved downstream. The results show the applicability of CARS to the non-intrusive concentration measurement of gas flows without any preprocess such as gas absorption into liquid or solid.

# 1. Introduction

Development of the microelectromechanical systems (MEMS) in the last few decades opened up the possibility to scale down microfluidic devices that leads to faster and more efficient mixing and/or reaction along with less sample consumption due to the increased surface-to-volume ratio. The miniaturization also enables the devices to be multi-functionalized for carrying out all chemical operations such as mixing, reaction, extraction and separation at once. Such devices are called micro total analysis systems (μTAS) or lab-on-a-chip, and expected to be used in vast fields, *e.g.* biology, chemistry, environmentology, medical science and pharmaceutical sciences [1–5].

Among various microfluidic chips, gas-specialized devices, which are called micro gas analysis system (μGAS), are expected to be used for gas sampling, generation, detection and separation [5–9] and numerous studies were conducted to understand the micro-scale flow behavior for obtaining the higher performance. However, less attempts were devoted for gas flows due to the difficulty of the gas flow measurement in mini- and micro-scale, whereas requirements for the μGAS are as high as those for the liquid-based systems [10]. Conventionally, some popular methods for the gas velocity measurement were developed such as the hot-wire airflow meter [11, 12] and the laser Doppler velocimetry (LDV) [13, 14]. Moreover, the particle image velocimetry (PIV) [15, 16] and the particle tracking velocimetry (PTV) [17] were developed in the end of the twentieth century as imaging techniques were progressed. The progress also brought a benefit for the temperature and concentration measurements, and the laser induced fluorescence (LIF) was developed [18]. However, less studies applied the LIF to the mini- and micro-scale gas flow measurements because the applicable tracers are limited to vaporized diatomic molecules that emit fluorescence [19–21]. $I_2$ and NO are examples that satisfy such requirement, and by employing them, Fujimoto and Niimi [20] and Mori *et al.* [21] respectively visualized a supersonic free jet of argon gas and low density high speed flows. Another attempt was conducted by Ichiyanagi *et al.* [22]. They measured the amount of $CO_2$ gas absorption into water in micro-scale $CO_2$–water parallel flows by adding fluorescein sodium salt into water as the fluorescent dye. Although the progression of the absorption was successfully observed, the concentration in the gas phase could not measure with this technique.

As it is obvious from above studies, the difficulty in inserting the tracers or dyes limits the applicability of these techniques to the gas flow measurement and measurement techniques that does not employ the tracers or dyes are required. In addition to this, a probeless technique is preferable if it is applied to mini- or micro-scale measurements because the probe could influence the flow of such scale. One of possible techniques that fulfill such requirements is the nuclear magnetic resonance (NMR) [23–27]. By employing NMR, the flow velocity, distribution or dispersion of both gas and liquid can be measured [23] if the fluids to be measured are NMR-active or can be detected indirectly by contrast agents or other sensors [27]. In other cases, the infra-red spectroscopy [28, 29] and the Raman spectroscopy [30] are seemingly prospects. Especially, the latter technique may be more suitable for measurements in micro-scale flows because of a feature that the measurement would less influenced by channel wall materials surrounding the target flow in comparison with the former.

The Raman spectroscopy has an ability to sort out molecules by detecting the scattered light from a sample, whose wavelength difference from the incident light is specific for the molecular species [31]. However, the signal emitted by the Raman effect is essentially very weak and it is crucial especially in case of the gas flow measurement because the intensity depends on the number density of molecules. This drawback, however, can be cleared by applying the coherent anti-Stokes Raman scattering (CARS), which has also an ability to detect the molecular species, and whose signal intensity is $10^3$–$10^6$ times as high as that of the spontaneous Raman scattering [32, 33]. Although CARS microscopy has a disadvantage of the point measurement technique due to the necessity of the focusing of the incident beams to induce the non-linear-optics effect, it is employed for the temperature and pressure measurement of $H_2$ [34] and the temperature measurement in combustion fields due to its high spatiotemporal resolution [35–38].

In this study a non-intrusive concentration measurement of a gas mixing flow by means of the CARS microscopy is performed. Carbon dioxide ($CO_2$) and nitrogen ($N_2$) are mixed in a millimeter-scale rectangular channel at the same flow rate. After obtaining calibration curves, which describe relationships between the CARS signal intensity and $CO_2$/$N_2$ concentrations, intensities of the CARS signal from each gas are measured at ten different points in the channel and the progression of the mixing is observed.

## 2. Measurement principle

When two high-energy laser beams of frequencies $\nu_1$ and $\nu_2$ ($\nu_1 > \nu_2$) are radiated collinearly to the sample, the beams interact coherently and produce the strong scattered light of frequency $2\nu_1 - \nu_2$ [39]. In addition to this, the Raman scattering is resonantly enhanced if the frequency difference of the laser beams coincides with eigen frequency of the sample $\nu$. This third-order non-linear optical effect is called the coherent anti-Stokes Raman scattering (CARS) and its intensity is $10^3$–$10^6$ times as high as that of the spontaneous Raman scattering [32, 33]. With incident beams called the pump and Stokes beams, the CARS process is shown in figure 1 and the frequency of the CARS beam $\nu_{CARS}$ is expressed as

$$\nu_{CARS} = 2\nu_{Pump} - \nu_{Stokes}, \tag{1}$$

where $\nu_{Pump}$ and $\nu_{Stokes}$ denote the frequencies of the pump and Stokes beams, respectively. Equation (1) implies that CARS is applicable to the detection of a specific molecule as well as the "spontaneous" Raman scattering by tuning the frequency difference between the pump and Stokes beams to the unique frequencies of the molecule, *i.e.* $\nu_{Pump} - \nu_{Stokes} = \nu$. Moreover, it is also possible to measure the concentration of the molecule because the CARS intensity is proportional to the square of the number density of the molecule $N$ as expressed in equation (2).

$$I_{CARS} \propto N^2 I_{Pump}^2 I_{Stokes}, \tag{2}$$

where $I$ denotes the intensity and the subscripts indicate each beam.

## 3. Experimental setup and procedure

### 3.1. Measurement system

Because two high-power laser beams with different wavelengths are essentially required and the wavelength difference must be adjusted to the Raman shift of the material to be measured, a tunable light source is necessitated. In our system, both beams are tunable, thereby the system can detect the CARS beam out of the incident beams by employing bandpass filters and dichroic mirrors. Figure 2 depicts a schematic of the system. A pulsed laser beam oscillated from a Ti:Sapphire laser (Coherent Inc., Chameleon Vision II, 680–1080 nm, 3.0 W, 140 fs, 80 MHz) was set to a wavelength for the Stokes beam and the beam was separated into two beams by a beam splitter. While one beam was used as the Stokes beam, another beam was introduced into an optical parametric oscillator (Coherent Inc., Chameleon Mira OPO) and its wavelength was converted to that for the pump beam. The pulses of two beams were then spatiotemporally overlapped by a micromotion stage and a dichroic mirror (Semrock Inc., FF746-SDi01-25×36×3.0). After extracting beams of undesired wavelengths by a pair of a dichroic mirror (Semrock Inc., FF709-Di01-25×2.0-D) and a filter (Semrock Inc., FF01-794/160-25), the overlapped beam was focused on one point by an upper objective (Nikon Corp., 20×, NA = 0.75, WD = 1.00) and the CARS signal was generated from the focal point. The generated CARS signal was collected by an EM-CCD camera (Hamamatsu Photonics K. K., C9100-12, 512 × 512 pixels, 16 × 16 μm$^2$/pixel, 14 bits) through a lower objective (Nikon Corp., 40×, NA = 0.6, WD = 3.6–2.8 mm) after filtering the incident beams by two filter blocks, which contain a pair of a dichroic mirror (Semrock Inc., FF518-Di01-25×36) and an optical filter (Semrock Inc., FF-01-650/60-25).

*3.2. Mixing channel*

Figure 3(*a*) shows a schematic of the rectangular test channel made of acrylic. Two gases were separately introduced through different inlets and mixed at the mixing region, and the concentration of two gases were measured at ten different points indicated in figure 3(*b*). The total length of the channel is 400 mm and the gas-introduction region (upstream of the mixing region) has a length of 100 mm. The cross-sectional dimension of the mixing region is 5 mm (height) × 35 mm (width) and hence the hydraulic diameter $d_\mathrm{h}$ is 8.75 mm. Each measurement point has a pair of hatches (open-mouth diameter: 13.8 mm) on the top and bottom walls of the test channel. The hatches were sealed during the measurement unless their location was of interest.

*3.3. Experimental conditions and procedure*

Carbon dioxide ($CO_2$) and nitrogen ($N_2$) were chosen as the test fluids. In consideration of the system modification such as the wavelength tuning and filtering, the combination of the wavelength of the pump and Stokes beams was determined based on the Raman shift (1388 cm$^{-1}$ for $CO_2$ and 2331 cm$^{-1}$ for $N_2$) as 720 nm and 800 nm for the $CO_2$ measurement and 727 nm and 876 nm for the $N_2$ measurement. The resulting CARS wavelengths were 655 nm ($CO_2$) and 622 nm ($N_2$), respectively.

Both gases were controlled by two float type flow meters (for $CO_2$: Tokyo Keiso Co., Ltd., P-100, 3–30 L/min, ±5% FS, for $N_2$: Tokyo Keiso Co., Ltd., P-200, 5–50 L/min, ±3% FS) and the flow rates were set to 10 L/min at 0.2 MPa (gauge), 20°C. The corresponding Reynolds numbers $Re$ were approximately 2200 ($CO_2$) and 1200 ($N_2$) at the gas-introduction region, and approximately 1700 at the mixing region, which was calculated with the averaged gas properties.

Before the measurement of the mixing, relationships between the concentration and CARS intensity (calibration curve) for the respective gases were obtained by a calibration experiment. The procedure of the calibration experiment was as follows: preliminarily mixed gas (*i.e.* known concentration) was introduced from one inlet with a flow rate of 10 L/min, which was controlled by the third flow meter (Tokyo Keiso Co., Ltd., P-200, 5–50 L/min, ±3%

FS), and the CARS intensity from either $CO_2$ or $N_2$ was measured at the measurement points. The concentrations of the gases were obtained afterwards by applying the calibration curves to the measured CARS intensities.

The exposure time of the EM-CCD camera was set to 1.57 second and five-successive images were averaged for one measurement. Note that the measurement was carried out by opening the pair of hatches on the channel wall and by inserting the objectives into the openings. This was necessary in order for the objectives to face each other at the focal point (the distance between them is approximately 10 mm) and in order to avoid the non-resonant background scattering whose wavelength is the same as the CARS signal.

## 4. Results and discussion

### 4.1. Calibration

A calibration experiment was first performed for the verification of the relationships between the concentration of each gas and CARS intensity. The $N_2/CO_2$ concentration ratio was adjusted to 0/100, 25/75, 50/50, 75/25 and 100/0 by the flow meters and the mixed gas was introduced from one inlet at 10 L/min, while another inlet was sealed. Figure 4 represents the relationships for $CO_2$ and $N_2$ obtained at five measurement points aligned in *X*-direction. Although the CARS intensity increased with the square of the concentration as it was expected from equation 2, the measured CARS intensities generated from the same concentration ratio did not always take single value. The discrepancy of the calibration curves could be due to the displacement of the distance between the objectives because the CARS intensity decreases by half if the focal point of the lower objective is displaced approximately 12 µm from the focal point of the upper objective as it is shown in figure 5. Therefore, in the following measurement, a calibration curve was newly obtained before measuring when the measurement point was changed.

### 4.2. Concentration measurement

$CO_2$ and $N_2$ were introduced to the test channel with respective flow rates of 10 L/min, and the intensity of the CARS signal was measured at ten different measurement points. Figures 6 and 7 show the obtained concentration distribution of $CO_2$ and $N_2$ in *X*-direction on $Y = \pm 9.25$ lines, respectively. As it is clear from figure 6, the $CO_2$ concentration at (0, −9.25) was approximately 100% and it decreased along $Y = −9.25$ line as the mixing with $N_2$ progressed. On the other hand, the $CO_2$ concentration increased from 0% to approximately 15% in 200 mm on $Y = +9.25$ line. The standard deviation at single measurement point was less than 3.90%. The same tendency was observed in case of the $N_2$ measurement (figure 7). In this case the increase in the concentration was found on $Y = -9.25$ line and the concentration decreased on $Y = +9.25$ line as the mixing progressed, and the standard deviation at single measurement point was less than 5.58%.

Because the Reynolds number was below 2300, the flow can be regarded as a laminar flow. Under this condition with an assumption that two gases flow in parallel, the mixing of gases in $Y$-direction was considered to progress by the diffusion because the pressure gradient in $Y$-direction, which was estimated by the calculated pressure drops of two fluids in the inlet region, was negligibly small (~0.33 Pa). On the other hand, it was assumed the effect of the diffusion in $X$-direction was negligible because the Peclet number $Pe$ expressed by equation (3) was ~$10^3$.

$$Pe = \frac{UL}{D}, \qquad (3)$$

where $U$, $L$ and $D$ (= $1.65 \times 10^{-5}$ m$^2$/s [40]) denote the flow velocity, the characteristic length and the diffusion coefficient for the $N_2$–$CO_2$ system, respectively. Note that the characteristic length of the diffusion $L_d$, which is known as the diffusion length, was calculated as ~$10^1$ μm from equation (4) with a timescale of $10^0$ second:

$$L_d = 2\sqrt{Dt}, \qquad (4)$$

where $t$ denotes time. Following the above result the residence time of molecules $t_m$ in the measurement section (from $X = 0$ mm to $X = 200$ mm) was estimated as

$$t_m = X/U. \qquad (5)$$

With this residence time it was possible to estimate the effect of the $Y$-directional diffusion in laminar flow and a one-dimensional and time-dependent diffusion analysis was performed with the initial condition that the concentration distribution of a gas is $n_0$ ($Y \geq 0$) and 0 ($Y < 0$). The concentration at arbitral position $Y$ can be calculated by the Fick's second law as a function of time described as equation (6).

$$n(Y, t_m) = \frac{1}{2} n_0 \, \text{erfc}\left(\frac{Y}{2\sqrt{Dt_m}}\right). \qquad (6)$$

Although the diffusion coefficient is a pressure-dependent [40], it was assumed that the pressure is approximately the same as the atmospheric because of the small pressure drop both in the channel and in the outlet jet [41]. With the initial condition that the residence time of molecules $t_m$ is 1 ms for $X = 0$ mm, the concentration distributions in $Y$-direction at $X = 0$ mm, 50 mm, 100 mm, 150 mm and 200 mm for $CO_2$ and $N_2$ were estimated and the results were represented in figure 8a and c, respectively. The results show the mixing progressed three times as significant as expected on a basis of $X$ and $t$, and the best fitting of the diffusion coefficient for equation (6) was derived as $D \sim 10^{-4}$ m$^2$/s (figure 8b and d). These results imply that a part of the mass transfer was due to the convection, namely the expansion of the channel path at the junction produced a weak but non-zero velocity in $Y$-direction. Furthermore, a difference in the flow development between the $CO_2$ and $N_2$ flows due to their different $Re$ might also influence. The entrance length $L_{ent}$ can be estimated by a correlation [42]

$$L_{ent} = 0.05 Re \cdot d_h, \qquad (7)$$

and $L_{ent}$ for $CO_2$ and $N_2$ flows were estimated as 832 mm and 449 mm, respectively. Because the flow profile changes ay every location along the entrance region, this difference in $L_{ent}$ could generate some turbulence resulted from the velocity difference of two fluids at the interface.

The results show that the CARS microscopy is applicable to the gas flow measurement as a direct non-intrusive measurement technique. Although the present measurement was carried out by removing the channel walls in order to avoid the non-resonant scattering from them, the measurement across the walls will be performed in future work by modifying angles of the incident beams to extract the non-resonant scattering. The measurement in microscale will be also achieved.

## 5. Concluding remarks

A direct and non-intrusive concentration measurement of a gas flow was performed by composing a CARS microscope system with a wavelength-tunable femtosecond pulsed laser, an optical parametric oscillator (OPO) and an EM-CCD camera. The system was applied to a concentration measurement in a millimeter-scale gas mixing flow. Main conclusions drawn from this study were summarized below.

With respect to the generality of the system, it was demonstrated that the tuning of the wavelengths of incident beams enabled the system to detect CARS signals from carbon dioxide ($CO_2$) and nitrogen ($N_2$) with common settings including optical filters. Furthermore, a concentration measurement in a gas mixing flow was carried out with the built-up system and the applicability of the CARS microscopy to the gas flow was proved.

The gas concentrations in a millimeter-scale rectangular channel were measured by employing a relationship between the $CO_2/N_2$ ratio and CARS intensity obtained by a calibration experiment. Both $CO_2$ and $N_2$ concentrations were measured at ten different points in the channel in which two gases were evenly introduced by a flow rate of 10 L/min.

Consequently, it was observed that the mixing progressed as the measurement point moved downstream. A simple analysis of the diffusion was also performed and it was shown that the diffusion coefficient derived from the experiment was one order as high as the conventional value. It was considered as the result of the turbulence generated by the expansion of the channel cross-section and the difference in the entrance length between two fluids.

Although the present measurement was conducted under the condition that the channel walls were physically extracted in order to avoid the negative influence caused by the non-resonant scattering, measurements across the channel walls will be performed in future work by modifying the incident angles of the pump and Stokes beams for suppressing the non-resonant scattering.

**Acknowledgements**

This work was supported by a Grant-in-Aid for Scientific Research (A) (No. 23246037), a Grant-in-Aid for Challenging Exploratory Research (No. 24656144), and a Grant-in-Aid for JSPS Fellows (No. 24–7201) (for the fourth author) from the Japan Society for the Promotion of Science and Keio Gijuku Academic Development Funds.

**References**

[1] Chin CD, Laksanasopin T, Cheung YK, Steinmiller D, Linder V, Parsa H, Wang J, Moore H, Rouse R, Umviligihozo G, Karita E, Mwambarangwe L, Braunstein SL, van de Wijgert J, Sahabo R, Justman JE, El-Sadr W and Sia SK 2011 Microfluidics-based diagnostics of infectious diseases in the developing world *Nat. Med.* **17** 1015–20

[2] Abate AR and Weitz DA 2011 Syringe-vacuum microfluidics: A portable technique to create monodisperse emulsions *Biomicrofluidics* **5** 014107

[3] Kreutzer MT, Kapteijn F, Moulijn JA and Heiszwolf JJ 2005 Multiphase monolith reactors: chemical reaction engineering of segmented flow in microchannels *Chem. Eng. Sci.* **60** 5895–916

[4] Liu B-F, Xu B, Zhang G, Du W and Luo Q 2006 Micro-separation toward systems biology *J. Chrom. A* **1106** 19–28

[5] Li S, Day JC, Park JJ, Cadou CP and Ghodssi R 2007 A fast-response microfluidic gas concentrating device for environmental sensing *Sens. Actuators A* **136** 69–79

[6] Ohira S and Toda K 2005 Micro gas analysis system for measurement of atmospheric

hydrogen sulfide and sulfur dioxide *Lab Chip* **5** 1374–9

[7] Ohira S, Someya K and Toda K 2007 In situ gas generation for micro gas analysis system *Anal. Chim. Acta* **588** 147–52

[8] Toda K, Hato Y, Ohira S and Namihira T 2007 Micro-gas analysis system for measurement of nitric oxide and nitrogen dioxide: Respiratory treatment and environmental mobile monitoring *Anal. Chim. Acta* **603** 60–6

[9] Hiki S, Mawatari K, Aota A, Saito M and Kitamori T 2011 Sensitive gas analysis system on a microchip and application for on-site monitoring of $NH_3$ in a clean room *Anal. Chem.* **83** 5017–22

[10] Ohira S and Toda K 2008 Micro gas analyzers for environmental and medical applications *Anal. Chim. Acta* **619** 143–56

[11] King LV 1914 On the convection of heat from small cylinders in a stream of fluid: determination of the convection constants of small platinum wires with applications to hot-wire anemometry *Phil. Trans. R. Soc. A* **214** 373–432

[12] Comte-Bellot G 1976 Hot-wire anemometry *Annu. Rev. Fluid Mech.* **8** 209–31

[13] Durst F, Melling A and Whitelaw JH 1976 Principles and practice of laser-Doppler anemometry (London: Academic Press)

[14] Albrecht H-E, Borys M, Damaschke N and Tropea C 2003 Laser Doppler and phase Doppler measurement techniques (Berlin: Springer)

[15] Adrian RJ 1991 Particle-imaging techniques for experimental fluid mechanics *Annu. Rev. Fluid Mech.* **23** 261–304

[16] Santiago JG, Wereley ST, Meinhart CD, Beebe DJ and Adrian RJ 1998 A particle image velocimetry system for microfluidics *Exp. Fluids* **25** 316–9

[17] Maas HG, Gruen A and Papantoniou D 1993 Particle tracking velocimetry in three-dimensional flows *Exp. Fluids* **15** 133–46

[18] Kinsey JL 1977 Laser-induced fluorescence *Annu. Rev. Phys. Chem.* **28** 349–72

[19] Niimi T 2013 Chapter 3: High Knudsen number flow ― optical diagnostic techniques, in *Micro-nano mechatronics ― new trends in material, measurement, control, manufacturing and their applications in biomedical engineering (Fukuda T, Niimi T and Obinata G ed.)* (Rijeka: InTech)

[20] Fujimoto T and Niimi T 1989 Three-dimensional structures of interacting freejets *Rarefied Gas Dynamics: Space-Related Studies (AIAA)* **116** 391–406

[21] Mori H, Niimi T, Taniguchi M, Nishihira R and Fukushima A 2005 Experimental analyses of linear-type aerospike nozzles with sidewalls *21st Int. Cong. Inst. Aerospace Sim. Fac.* 145–9

[22] Ichiyanagi M, Tsutsui I, Kakinuma Y, Sato Y and Hishida K 2012 Three-dimensional

measurement of gas dissolution process in gas–liquid microchannel flow *Int. J. Heat Mass Transfer* **55** 2872–8

[23] Zhivonitko VV, Telkki V-V and Koptyug IV 2012 Characterization of microfluidic gas reactors using remote-detection MRI and parahydrogen-induced polarization *Angew. Chem.* **51** 8054–8

[24] Zhivonitko VV, Telkki V-V, Leppäniemi J, Scotti G, Franssila S and Koptyug IV 2013 Remote detection NMR imaging of gas phase hydrogenation in microfluidic chips *Lab Chip* **13** 1554–61

[25] Hilty C, McDonnell EE, Granwehr J, Pierce KL, Han S-I and Pines A 2005 Microfluidic gas-flow profiling using remote-detection NMR *PNAS* **102** 14960–3

[26] Harel E, Hilty C, Koen K, McDonnell EE and Pines A 2007 Time-of-flight flow imaging of two-component flow inside a microfluidic chip *Phys. Rev. Lett.* **98** 017601

[27] Bajaj VS, Paulsen J, Harel E and Pines A 2010 Zooming in on microscopic flow by remotely detected MRI *Science* **330** 1078–81

[28] McCurdy MR, Bakhirkin YA and Tittel FK 2006 Quantum cascade laser-based integrated cavity output spectroscopy of exhaled nitric oxide *Appl. Phys. B* **85** 445–52

[29] Kosterev A, Wysocki G, Bakhirkin Y, So S, Lewicki R, Fraser M, Tittel F and Curl RF 2008 Application of quantum cascade lasers to trace gas analysis *Appl. Phys. B* **90** 165–76

[30] Welsh HL, Crawford MF, Thomas TR and Love GR 1952 Raman spectroscopy of low pressure gases and vapors *Can. J. Phys.* **30** 577–96

[31] Raman CV and Krishnan KS 1928 A new type of secondary radiation *Nature* **121** 501–2

[32] Maker PD and Terhune RW 1965 Study of optical effects due to an induced polarization third order in the electric field strength *Phys. Rev.* **137** A801–18

[33] Begley RF, Harvey AB and Byer RL 1974 Coherent antiStokes Raman spectroscopy *Appl. Phys. Lett.* **25** 387–90

[34] Lang T, Kompa K-L and Motzkus M 1999 Femtosecond CARS on $H_2$ *Chem. Phys. Lett.* **310** 65–72

[35] Brackmann C, Bood J, Afzelius M and Bengtsson P-E 2004 Thermometry in internal combustion engines via dual-broadband rotational coherent anti-Stokes Raman spectroscopy *Meas. Sci. Technol.* **15** R13–25

[36] Roy S, Meyer TR, Lucht RP, Belovich VM, Corporan E and Gord JR 2004 Temperature and $CO_2$ concentration measurements in the exhaust stream of a liquid-fueled combustor using dual-pump coherent anti-Stokes Raman scattering (CARS) spectroscopy *Combust. Flame* **138** 273–84

[37] Beyrau F, Bräuer A, Seeger T and Leipertz A 2004 Gas-phase temperature measurement in the vaporizing spray of a gasoline direct-injection injector by use of pure rotational

coherent anti-Stokes Raman scattering *Opt. Lett.* **29** 247–9

[38] Bohin A and Kliewer CJ 2014 Diagnostic imaging in flames with instantaneous planar coherent Raman spectroscopy *J. Phys. Chem. Lett.* **5** 1243–8

[39] Ferraro JR, Nakamoto K and Brown CW 2003 Introductory Raman spectroscopy (Amsterdam: Elsevier)

[40] Marrero TR and Mason EA 1972 Gaseous diffusion coefficients *J. Phys. Chem. Ref. Data* **1** 3–118

[41] Rajaratnam N 1976 Turbulent jets (Amsterdam: Elsevier)

[42] Kandlikar SG, Garimella S, Li D, Colin S and King MR 2006 Heat transfer and fluid flow in minichannels and microchannels (Amsterdam: Elsevier)

**FIGURES**

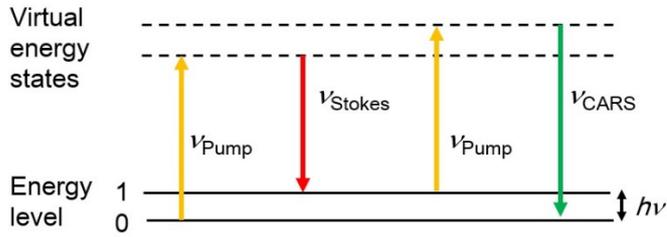

**Figure 1.** Photon process of the CARS (color online).

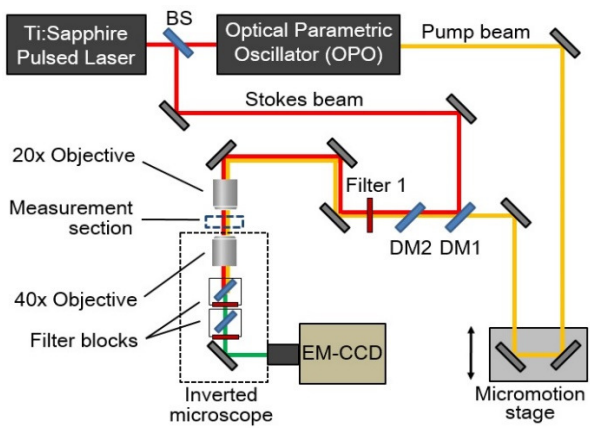

**Figure 2.** A schematic of the CARS microscope system equipped for the present study (color online).

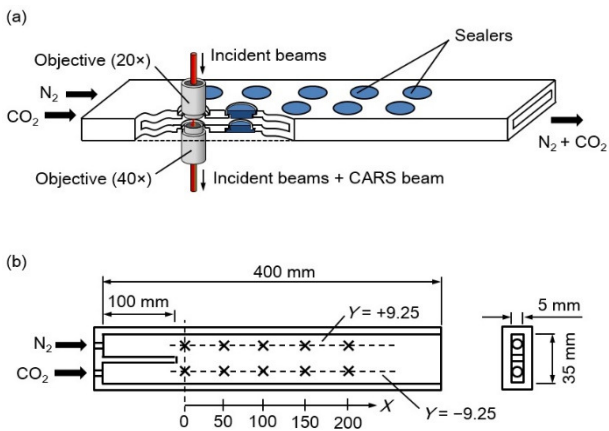

**Figure 3.** (*a*) A schematic of the test channel. (*b*) Top and cross-sectional views of the test channel with dimensions. The measurement points are aligned in two parallel lines at intervals of 50 mm (color online).

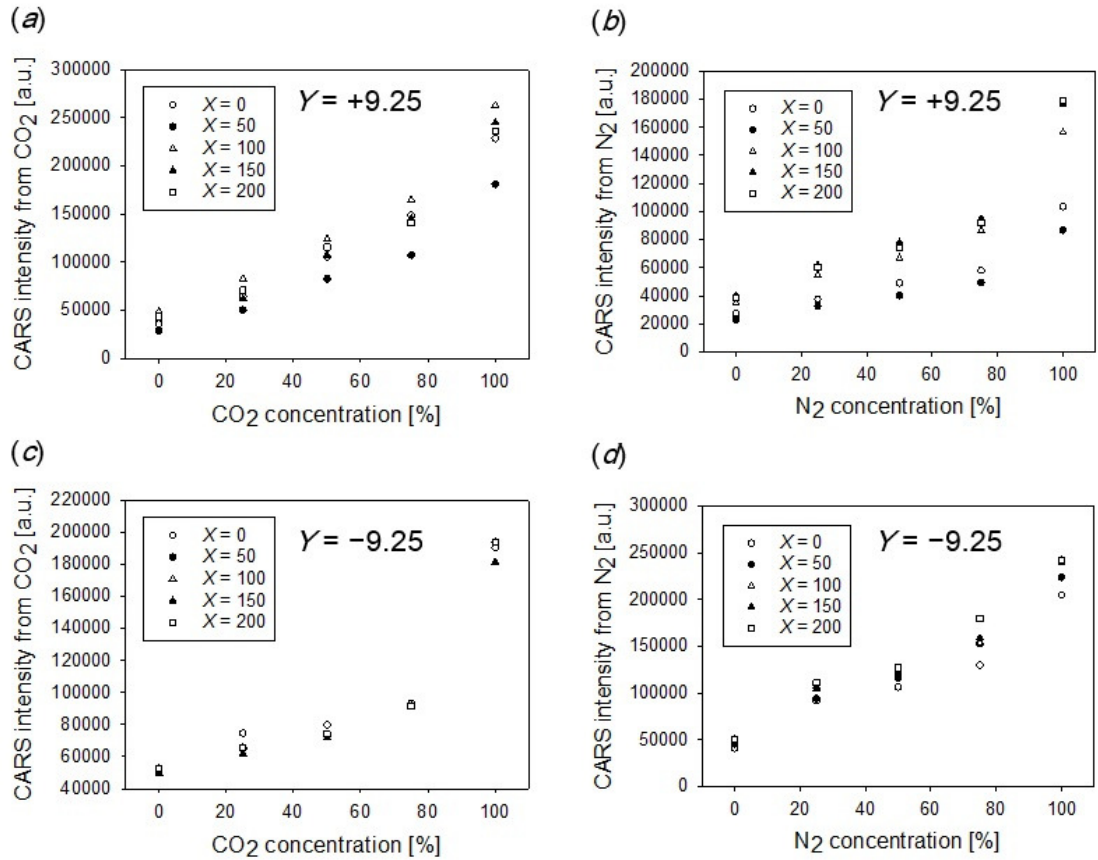

**Figure 4.** Calibration curves obtained at five measurement points on single line for (*a*) $CO_2$ at $Y$ = +9.25, (*b*) $N_2$ at $Y$ = +9.25, (*c*) $CO_2$ at $Y$ = −9.25, (*d*) $N_2$ at $Y$ = −9.25. The figures indicate that the CARS intensity is proportional to the square of the concentration, but the relationships cannot be expressed by a universal correlation.

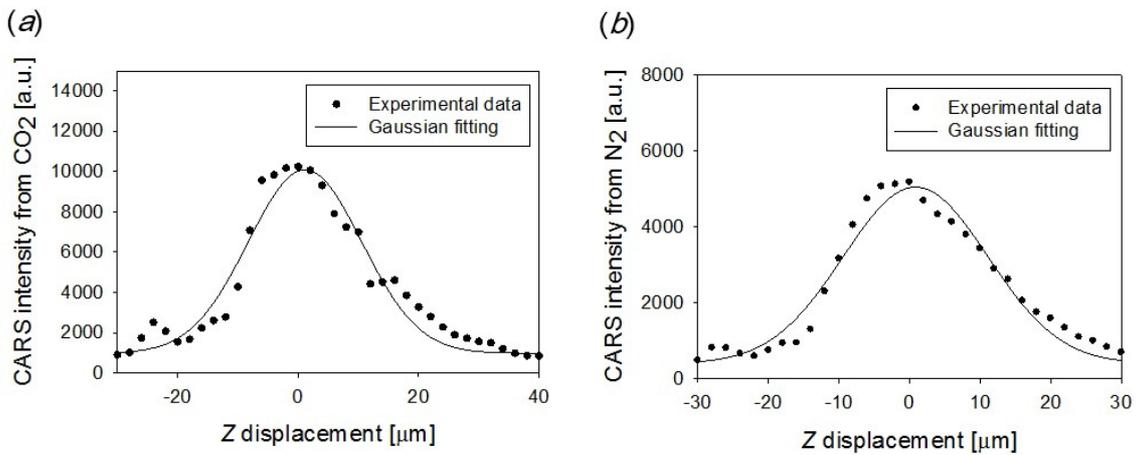

**Figure 5.** CARS intensity distribution in $Z$-direction of (*a*) carbon dioxide and (*b*) nitrogen. The FWHM (full width at half maximum) of both signals is approximately 24 μm.

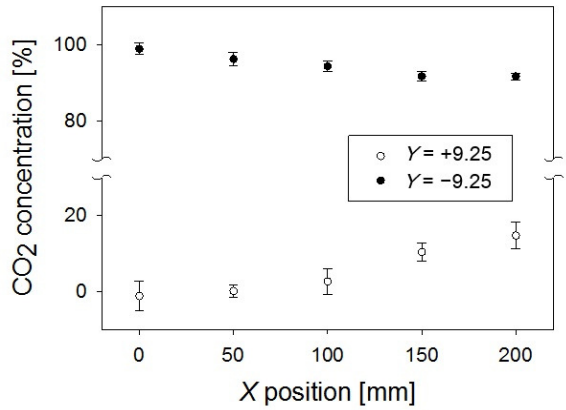

**Figure 6.** $CO_2$ concentration distribution in *X*-direction on $Y = \pm 9.25$ lines. The maximum standard deviation is 3.90% at $X = 0$ mm.

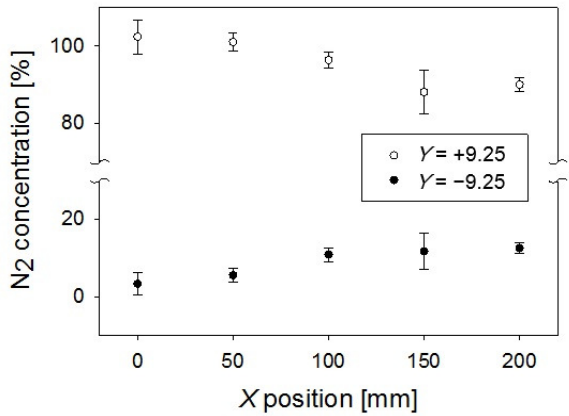

**Figure 7.** $N_2$ concentration distribution in *X*-direction on $Y = \pm 9.25$ lines. The maximum standard deviation is 5.58% at $X = 150$ mm.

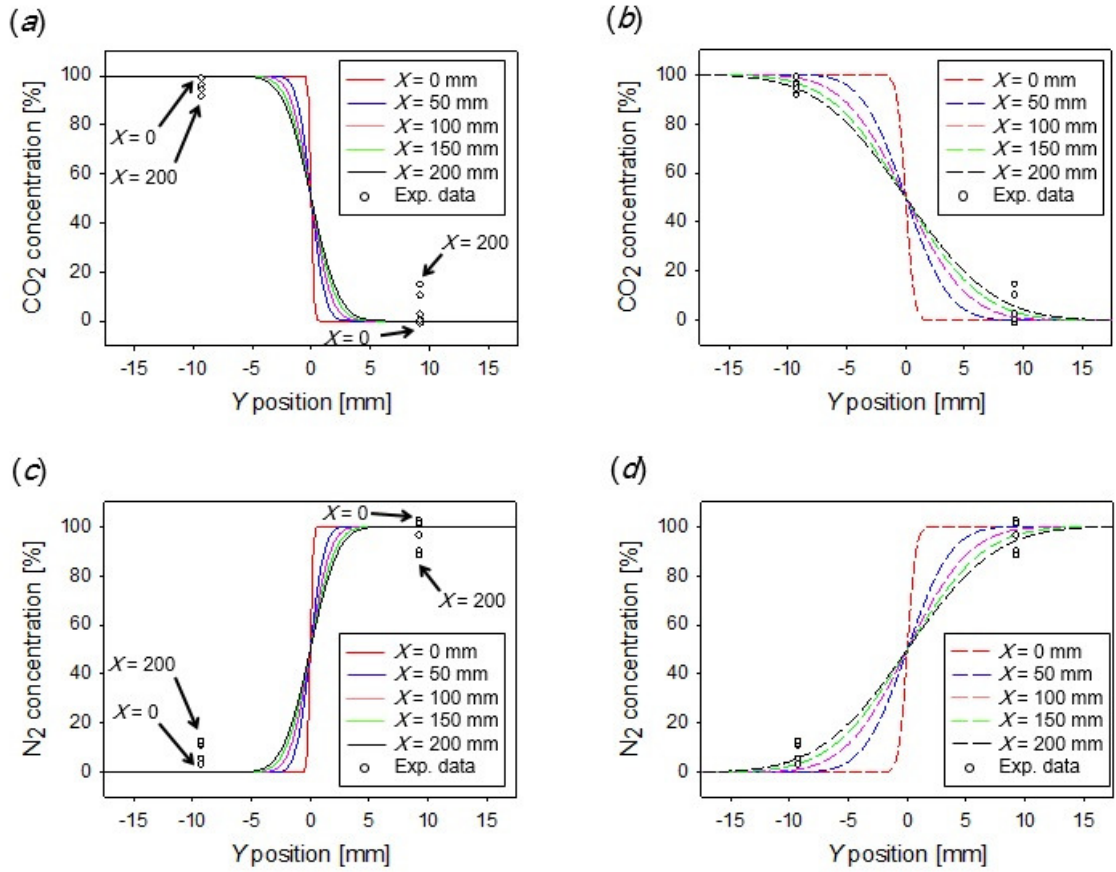

**Figure 8.** Theoretical concentration distributions in *Y*-direction at *X* = 0 mm (red), 50 mm (blue), 100 mm (pink), 150 mm (green) and 200 mm (black) for (*a*) $CO_2$ calculated with $D = 1.65 \times 10^{-5}$ m²/s, (*b*) $CO_2$ calculated with $D = 1.65 \times 10^{-4}$ m²/s, (*c*), $N_2$ calculated with $D = 1.65 \times 10^{-5}$ m²/s, and (*d*) $N_2$ calculated with $D = 1.65 \times 10^{-4}$ m²/s. Measured concentrations at $Y = \pm 9.25$ mm are represented with circles (color online).